\documentclass[aps,prl,twocolumn,showpacs,preprintnumbers,amsmath,amssymb,refmerge,superscriptaddress]{revtex4}
\usepackage{graphicx}
\usepackage{dcolumn}
\usepackage{longtable}
\begin{document}

\title{Observation of New States Decaying into $\Lambda_c^+ K^-\pi^+$ and $\Lambda_c^+ K^0_S\pi^-$}

\affiliation{Budker Institute of Nuclear Physics, Novosibirsk}
\affiliation{Chiba University, Chiba}
\affiliation{Chonnam National University, Kwangju}
\affiliation{University of Cincinnati, Cincinnati, Ohio 45221}
\affiliation{University of Hawaii, Honolulu, Hawaii 96822}
\affiliation{High Energy Accelerator Research Organization (KEK), Tsukuba}
\affiliation{University of Illinois at Urbana-Champaign, Urbana, Illinois 61801}
\affiliation{Institute of High Energy Physics, Chinese Academy of Sciences, Beijing}
\affiliation{Institute of High Energy Physics, Vienna}
\affiliation{Institute of High Energy Physics, Protvino}
\affiliation{Institute for Theoretical and Experimental Physics, Moscow}
\affiliation{J. Stefan Institute, Ljubljana}
\affiliation{Kanagawa University, Yokohama}
\affiliation{Korea University, Seoul}
\affiliation{Kyungpook National University, Taegu}
\affiliation{Swiss Federal Institute of Technology of Lausanne, EPFL, Lausanne}
\affiliation{University of Ljubljana, Ljubljana}
\affiliation{University of Maribor, Maribor}
\affiliation{University of Melbourne, Victoria}
\affiliation{Nagoya University, Nagoya}
\affiliation{Nara Women's University, Nara}
\affiliation{National Central University, Chung-li}
\affiliation{National United University, Miao Li}
\affiliation{Department of Physics, National Taiwan University, Taipei}
\affiliation{H. Niewodniczanski Institute of Nuclear Physics, Krakow}
\affiliation{Nippon Dental University, Niigata}
\affiliation{Niigata University, Niigata}
\affiliation{University of Nova Gorica, Nova Gorica}
\affiliation{Osaka City University, Osaka}
\affiliation{Osaka University, Osaka}
\affiliation{Panjab University, Chandigarh}
\affiliation{Peking University, Beijing}
\affiliation{Princeton University, Princeton, New Jersey 08544}
\affiliation{RIKEN BNL Research Center, Upton, New York 11973}
\affiliation{Saga University, Saga}
\affiliation{University of Science and Technology of China, Hefei}
\affiliation{Seoul National University, Seoul}
\affiliation{Sungkyunkwan University, Suwon}
\affiliation{University of Sydney, Sydney NSW}
\affiliation{Tata Institute of Fundamental Research, Bombay}
\affiliation{Toho University, Funabashi}
\affiliation{Tohoku Gakuin University, Tagajo}
\affiliation{Tohoku University, Sendai}
\affiliation{Department of Physics, University of Tokyo, Tokyo}
\affiliation{Tokyo Institute of Technology, Tokyo}
\affiliation{Tokyo Metropolitan University, Tokyo}
\affiliation{Tokyo University of Agriculture and Technology, Tokyo}
\affiliation{Toyama National College of Maritime Technology, Toyama}
\affiliation{University of Tsukuba, Tsukuba}
\affiliation{Virginia Polytechnic Institute and State University, Blacksburg, Virginia 24061}
\affiliation{Yonsei University, Seoul}
 \author{R.~Chistov}\affiliation{Institute for Theoretical and Experimental Physics, Moscow} 
  \author{K.~Abe}\affiliation{High Energy Accelerator Research Organization (KEK), Tsukuba} 
  \author{K.~Abe}\affiliation{Tohoku Gakuin University, Tagajo} 
  \author{I.~Adachi}\affiliation{High Energy Accelerator Research Organization (KEK), Tsukuba} 
  \author{H.~Aihara}\affiliation{Department of Physics, University of Tokyo, Tokyo} 
  \author{D.~Anipko}\affiliation{Budker Institute of Nuclear Physics, Novosibirsk} 
  \author{V.~Aulchenko}\affiliation{Budker Institute of Nuclear Physics, Novosibirsk} 
  \author{T.~Aushev}\affiliation{Institute for Theoretical and Experimental Physics, Moscow} 
  \author{A.~M.~Bakich}\affiliation{University of Sydney, Sydney NSW} 
  \author{V.~Balagura}\affiliation{Institute for Theoretical and Experimental Physics, Moscow} 
% ...
\collaboration{The Belle Collaboration}

\begin{abstract}

We report the first observation of two charmed strange baryons that decay into 
$\Lambda_c^+ K^-\pi^+$. 
The broader of the two states is measured to have a mass of 
$2978.5\pm 2.1\pm 2.0$~MeV/$c^2$ and a width of 
$43.5\pm 7.5\pm 7.0$~MeV/$c^2$. 
The mass and width of the narrow state are measured
to be $3076.7\pm 0.9\pm 0.5$~MeV/$c^2$  and $6.2\pm 1.2\pm 0.8$~MeV/$c^2$, 
respectively. 
We also perform a search for the isospin partner states that decay into 
$\Lambda_c^+ K_S^0\pi^-$ and observe a 
significant signal at the mass of $3082.8\pm 1.8\pm 1.5$~MeV/$c^2$.
The data used for this analysis was accumulated at or near 
the $\Upsilon(4S)$ resonance, using the Belle detector at the 
$e^+ e^-$ asymmetric-energy collider KEKB.
The integrated luminosity of the data sample used is   
$461.5\,~\mathrm{fb}^{-1}$.

\end{abstract}


\maketitle

{\renewcommand{\thefootnote}{\fnsymbol{footnote}}}
\setcounter{footnote}{0}

Several excited $\Lambda_c^+$~\cite{lmc1}, $\Sigma_c$~\cite{sigmac} and $\Xi_c$~\cite{new_xic}
baryons have already been observed. 
The most recent examples are an isotriplet of excited 
$\Sigma_c$ baryons 
and the $\Lambda_c(2940)^+$, reported by Belle and BaBar, respectively \cite{sigmac2800}. 
The charmed baryon sector offers a rich source of states and possible 
orbital excitations, serving as an excellent laboratory to test 
the predictions of the quark model and other models of bound quarks~\cite{theory_one}, as 
well as predictions based on heavy quark symmetry~\cite{theory_two}.
Some of the recently discovered baryons are candidates for 
orbitally excited states~\cite{pdg}.
Within the $\Xi_c$ system, two candidates for the first P-wave orbital excitations were found, the $\Xi_c(2790)$ and $\Xi_c(2815)$ baryons, which decay into $\Xi_c'\pi$ and $\Xi_c^*\pi$, respectively~\cite{xic12, xic11}. 
In these decays, the charm and strange quarks of the initial state are
inherited by the final state baryon.
However, nothing is experimentally known about charmed strange 
baryons that decay to $\Lambda_c^+ K^-\pi^+$. 
In such a decay processes, the charm and strangeness of the initial state 
are carried away by two different final state particles, a charmed baryon 
and a strange meson.

The SELEX collaboration has reported the observation of a doubly 
charmed baryon that decays into the $\Lambda_c^+ K^-\pi^+$ final state~\cite{selex}. 
The SELEX claim has not been confirmed by other experiments. 
Combined with the 
measured cross section for double $c\bar{c}$ production~\cite{additional}, 
which is an order of magnitude larger than non-relativistic 
QCD predictions~\cite{additional2}, 
a search for the SELEX doubly charmed baryon is an additional motivation 
for the examination of the $\Lambda_c^+ K^-\pi^+$ final state.

In this Letter we report the results of a search for new baryons 
decaying into $\Lambda_c^+ K^-\pi^+$ and $\Lambda_c^+ K^0_S\pi^-$ final states.
Inclusion of charge conjugate states is implicit unless otherwise stated.
The analysis is performed using data collected
with the Belle detector at the KEKB asymmetric-energy $e^+e^-$
collider~\cite{kekb}.  
The data sample corresponds to an integrated luminosity of 
$461.5$~fb$^{-1}$ collected at or near the
$\Upsilon(4S)$ resonance.

The Belle detector is a large-solid-angle magnetic spectrometer that consists of a silicon
vertex detector (SVD), a 50-layer central drift chamber (CDC), an array of aerogel threshold
Cherenkov counters (ACC), a barrel-like arrangement of time-of-flight scintillation counters
(TOF), and an electromagnetic calorimeter (ECL) comprised of CsI(Tl) crystals located
inside a superconducting solenoid coil that provides a 1.5 T magnetic field. An iron flux-return located outside of the coil is instrumented to detect $K^0_L$ mesons and to identify muons
(KLM). The detector is described in detail elsewhere~\cite{belle_detector}. 
Two different inner detector configurations 
 were used, a 2.0~cm beam-pipe and a 3-layer 
silicon vertex detector for the first 155~fb$^{-1}$, 
and a 1.5~cm beam-pipe with 
a 4-layer vertex detector for the remaining 306.5 ~fb$^{-1}$~\cite{svd2}.
We use a GEANT-based Monte Carlo (MC) simulation to model 
the response of the detector and determine the efficiency~\cite{montecarlo}.

Protons, charged pions and kaons are required to originate from the region,
$dr<1$~cm, $|dz|<3$~cm. Here, $dr$ and $dz$ are the distances of
closest approach to the interaction point in the plane perpendicular to the 
beam axis ($r-\phi$ plane) and along the beam direction, respectively.
Charged hadrons are identified using a likelihood ratio method, 
which combines information from the TOF system and ACC counters with $dE/dx$ measurements in the CDC. 
To identify charged particles ($\pi$, $K$, $p$), we apply the standard Belle requirements on the corresponding likelihood ratios~\cite{belle_detector}.
Neutral kaons are reconstructed via the decay $K_S^0\to\pi^+\pi^-$, requiring  
$M(\pi^+\pi^-)$ to be  
within $\pm 10$~MeV/$c^2$ of the nominal $K_S^0$ mass~\cite{pdg}.
We require the displacement of the $\pi^+\pi^-$ vertex from the interaction point in the $r-\phi$ plane to be more than 0.1~cm.

We reconstruct the $\Lambda_c^+$ via the $\Lambda_c^+\to pK^-\pi^+$ decay channel. 
All $pK^-\pi^+$ combinations with an invariant mass within $\pm 10$~MeV/$c^2$ ($\sim 2.5~\sigma$) around 2286.6~MeV/$c^2$ are selected as $\Lambda_c^+$ candidates. 
The mean value of the mass for our $\Lambda_c^+$ signal is 
$2286.6\pm 0.1~(stat.)$~MeV/$c^2$, 
in good agreement with a recent precision measurement by 
BaBar~\cite{babar_lamc_mass}.
We perform a mass constrained fit to the $\Lambda_c^+$ vertex and then 
combine $\Lambda_c^+$ candidates 
with the remaining $K^-\pi^+$ pairs in the event.

\begin{table*}[t]
\begin{ruledtabular}
\caption {Summary of the parameters of the 
new states in the $\Lambda_c^+ K^-\pi^+$ 
and $\Lambda_c^+ K_S^0\pi^+$ final states: 
masses, widths, yields and statistical significances.}
\begin{tabular}{ccccc}
New State  &  Mass  (MeV/$c^2$)  & Width  (MeV/$c^2$) & Yield  (events)  & Significance ($\sigma$) \\
\hline
$\Xi_{cx}(2980)^+$ & $2978.5\pm 2.1\pm 2.0$  &   $43.5\pm 7.5\pm 7.0$   & $405.3\pm 50.7$  & 6.3   \\
$\Xi_{cx}(3077)^+$ &  $3076.7\pm 0.9\pm 0.5$ & $6.2\pm 1.2\pm 0.8$  &  $326.0\pm 39.6$ & 9.7   \\
\hline
$\Xi_{cx}(2980)^0$ &    $2977.1\pm 8.8\pm 3.5$        & $43.5$ (fixed)         &   $42.3\pm 23.8$ &   2.0      \\
$\Xi_{cx}(3077)^0$ &    $3082.8\pm 1.8\pm 1.5$        & $5.2\pm 3.1\pm 1.8$        &    $67.1\pm 19.9$ & 5.1       \\
\end{tabular}
\label{table_sum_fit}
\end{ruledtabular}
\end{table*}

\begin{figure*}[!t]
\includegraphics[width=0.45\textwidth]{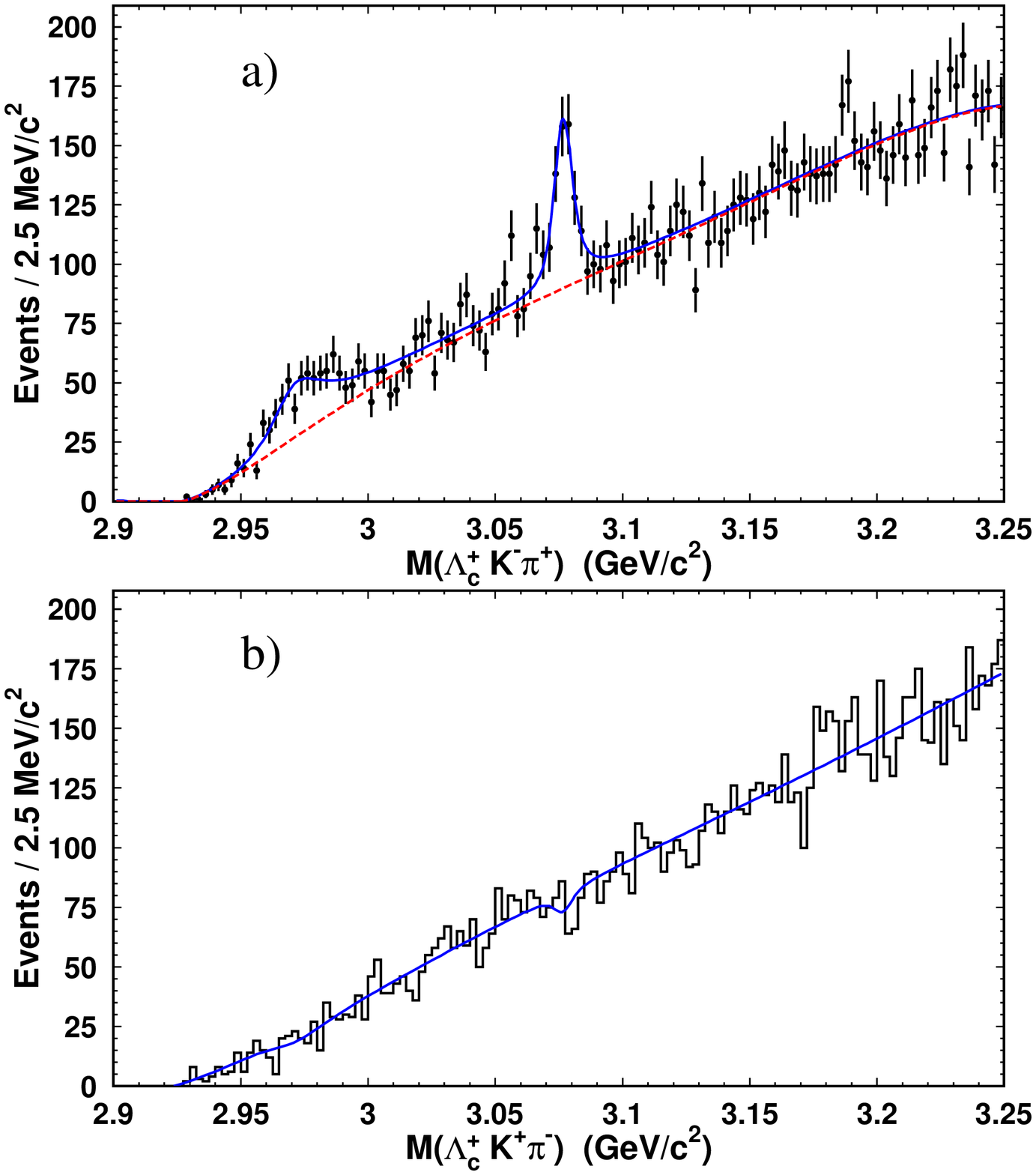}
\includegraphics[width=0.45\textwidth]{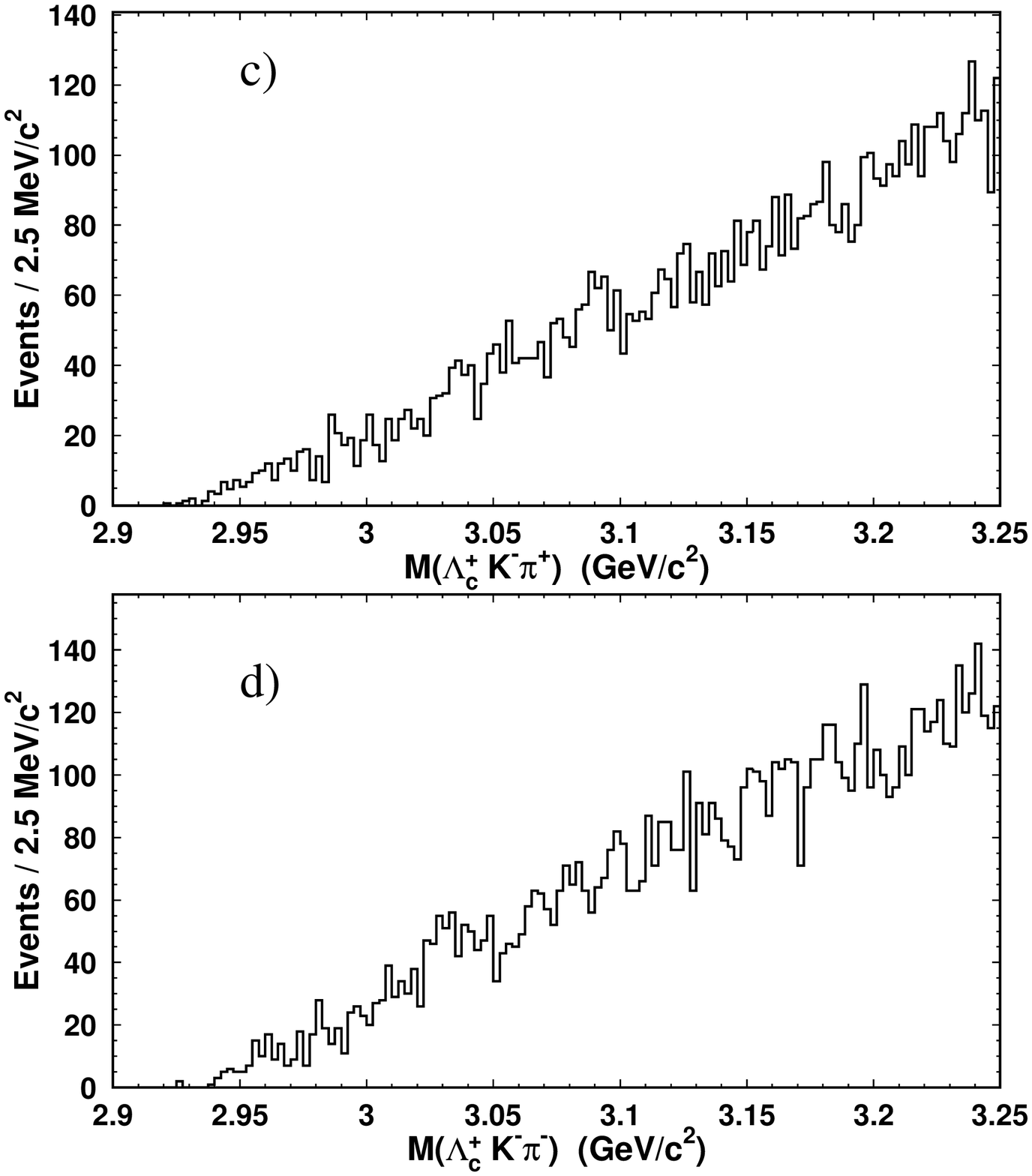}
 \caption{(a) $M(\Lambda_c^+ K^-\pi^+)$ distribution together with the overlaid fitting curve. 
Points with errors represent the data, dashed line is the background component of the
fitting function described in the text, and the solid curve  is the
sum of the background and signal. 
(b) The WS  
combination mass distribution $M(\Lambda_c^+ K^+\pi^-)$ fitted with the same 
function including the signal components where the masses and widths of the signals are fixed to the values from the fit to the RS distribution. 
Two additional cross-checks are shown (c) 
the invariant mass distribution of the right-sign $\Lambda_c^+ K^-\pi^+$ combinations but using appropriately scaled sidebands of the $\Lambda_c^+$ signal 
and (d) the invariant mass distribution of the other WS $\Lambda_c^+ K^-\pi^-$ combinations.  
No structures are visible in the signal regions near 2980~MeV/$c^2$ and 3077~MeV/$c^2$.  
}
  \label{f1}
\end{figure*}

The momentum spectra of charmed hadrons produced 
in $e^+e^-\to c\bar{c}$ continuum are hard compared to the combinatorial 
background. Therefore, we apply the requirement $p^* >3.0$~GeV/$c$, 
where $p^*$ is the momentum of the $\Lambda_c^+ K^- \pi^+$  system in 
the  center of mass frame.
We also fit the $\Lambda_c^+ K^-\pi^+$ combinations to a common vertex.

The resulting invariant mass distribution $M(\Lambda_c^+ K^-\pi^+)$ 
is shown in Fig.~\ref{f1}~(a). 
Two peaks are visible in this distribution: a broad one near 
threshold at a mass of about 2980 MeV/$c^2$ and a narrower one 
at a higher mass of about 3077 MeV/$c^2$.
We verify that the observed signals are robust and their mass 
values stable against the variation of particle identification criteria, 
$\Lambda_c^+$ mass selection window and the $p^*$ requirement.   
Hereafter we denote the observed peaks as $\Xi_{cx}(2980)^+$ and  $\Xi_{cx}(3077)^+$ as explained below.

Fig.~\ref{f1}~(b) shows the invariant mass distribution of the wrong-sign 
(WS) combinations $M(\Lambda_c^+ K^+ \pi^-)$, which has a smooth 
structureless behaviour. This demonstrates that the observed 
peaks in the right-sign (RS) $M(\Lambda_c^+ K^- \pi^+)$ invariant mass 
distribution are not reflections due to $K-\pi$ misidentification 
originating from the four known excited baryons
$\Lambda_c(2593)^+$, $\Lambda_c(2625)^+$, $\Lambda_c(2765)^+$ 
and $\Lambda_c(2880)^+$~\cite{lmc1, pdg} that 
decay into $\Lambda_c^+\pi^+\pi^-$.
Reflections from the $\Lambda_c^+ \pi^+ \pi^-$ decay modes of these 
states would contribute equally to both the RS and WS distributions.
This conclusion is also verified with a MC simulation of $\Lambda_c(2593)^+$, $\Lambda_c(2625)^+$, $\Lambda_c(2765)^+$ and $\Lambda_c(2880)^+$ produced in $e^+e^-\to c\bar{c}$.
We generate $10^4$ $\Lambda_c^+\pi^+\pi^-$ decays for each excited  
$\Lambda_c^+$ state and 
reconstruct the MC events with the same selection criteria as the data.
The  resulting mass distributions exhibit similar behaviour in both RS and WS cases. 
Simulation shows that the contribution of excited $\Lambda_c^+$ baryons
is reduced to $1.2\%$ with the above selection requirements. Using the
reconstruction of $\Lambda_c(2880)^+\to \Lambda_c^+ \pi^+ \pi^-$ 
on the same data set we find that possible 
contribution of reflections is below the statistical
sensitivity of our measurement.

\begin{figure}[!b]
\centering
\includegraphics[width=.45\textwidth]{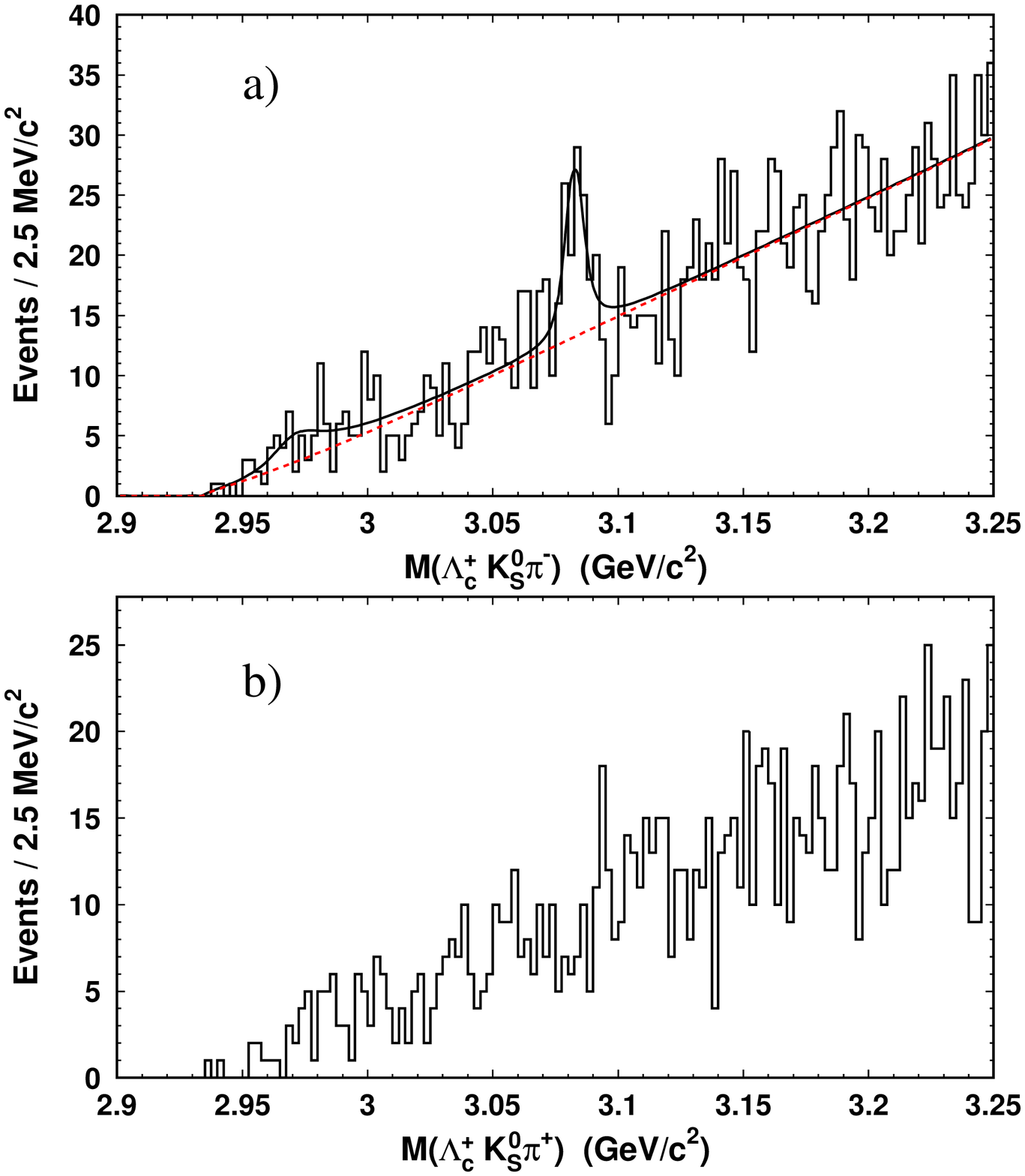}
\caption{ (a) $M(\Lambda_c^+ K^0_S\pi^-)$ distribution together with the overlaid fitting curve. 
The fitting function is the same as in the $\Lambda_c^+ K^+\pi^-$ case 
 (see the text).
(b) The WS 
combination mass distribution $M(\Lambda_c^+ K^0_S\pi^+)$.
}
\label{fks} 
\end{figure}

Fig.~\ref{f1}~(c) shows an additional check of the $M(\Lambda_c^+ K^-\pi^+)$ 
distribution of data events in the $\Lambda_c^+$ 
mass sidebands~\cite{lmc_sidebands}.
We also check the invariant mass distribution for the 
$\Lambda_c^+ K^-\pi^-$ WS combination,  
shown in 
Fig.~\ref{f1}~(d). Both distributions in Fig.~\ref{f1}~(c) and (d) 
are featureless near the 2980 MeV/$c^2$ and 3077 MeV/$c^2$ mass regions.

Results of the fit to the $M(\Lambda_c^+ K^-\pi^+)$ 
distribution are shown by the solid curve in Fig.~\ref{f1}~(a).
The simulated mass resolution of $\Xi_{cx}(2980)\to \Lambda_c^+ K^- \pi^+$ is found to be $1.2$~MeV/$c^2$,  which is 
much smaller than the observed signal width.
Therefore, the  broad signal near the threshold is modeled by 
a Breit-Wigner function only.
To describe the $\Xi_{cx}(3077)^+$ resonance we use a 
Breit-Wigner convolved with a Gaussian detector 
resolution function. The width of the Gaussian ($\sigma$) is fixed from 
MC to be 2.0~MeV/$c^2$. 
The background is described by a threshold function 
${\rm atan}(\sqrt{x - x_{\rm thr}})$
multiplied by a third-order polynomial.  
The dashed line in Fig.~\ref{f1}~(a) shows the background component of the fitting 
function. The results of the fit are summarised in  
Table~\ref{table_sum_fit}. 
The $\chi^2/ndf$ of the fit is $0.98$. 
The statistical significance of each of the two observed signals is defined as 
$\sqrt{-2{\rm ln}(L_0/L_{\rm max})}$. 
Here, $L_0$ and $L_{{\rm max}}$ are the values of the likelihood function with the corresponding signal
fixed to zero and at the best fit value, respectively.
We fit the WS mass distribution using the same parametrization, with 
parameters describing the shape of the signal fixed to the above values. 
The fit yields $-34.8\pm 19.6$ ($-78.2\pm 54.6$) 
events for the higher (lower) mass peak,
consistent with zero.

%
To provide more information on the origin of the states found in the present 
analysis,    
we perform 
a search for their neutral isospin partners in the $\Lambda_c^+ K^0_S\pi^-$ 
final state. 
In this case the selection criteria are the same as for the  
$\Lambda_c^+ K^-\pi^+$
final state with one exception: a tighter momentum requirement, 
$p^*>3.5$~GeV/$c$, is applied for the $\Lambda_c^+ K^0_S\pi^-$ system.
The resulting invariant mass distribution, $M(\Lambda_c^+ K^0_S\pi^-)$,  
is shown in Fig.~\ref{fks}~(a), where a clear signal near 3077~MeV/$c^2$ is observed.
A broad 
enhancement near the threshold can also be identified.  
Fig.~\ref{fks}~(b) shows the WS $M(\Lambda_c^+ K^0_S\pi^+)$ 
distribution, which is featureless.  
To describe the narrow signal, 
which we denote as $\Xi_{cx}(3077)^0$,  
we use a Breit-Wigner function 
convolved with a Gaussian detector resolution function. 
The width of the Gaussian is fixed from MC to be $\sigma=2.4$~MeV/$c^2$. 
To describe the broad signal near the threshold which we denote as $\Xi_{cx}(2980)^0$ we use a Breit-Wigner function with the width fixed to that of $\Xi_{cx}(2980)^+$, $\Gamma=43.5$~MeV/$c^2$. 
The background is described by a threshold function multiplied by a third-order polynomial.  
The signal yields and parameters of the two Breit-Wigner 
functions determined from the fit are given in Table~\ref{table_sum_fit}.

We also vary order of the polynomial for 
the background function and the widths 
of the detector resolution within their errors. The resulting 
systematic uncertainties are given in Table~\ref{table_sum_fit}. 
None of the variations reduces the significances of the 
$\Xi_{cx}(3077)^+$, $\Xi_{cx}(2980)^+$ and $\Xi_{cx}(3077)^0$ to less 
than 9$\sigma$, $6\sigma$ and $5\sigma$, respectively.

\begin{figure}[!b]
\centering
\includegraphics[width=0.45\textwidth]{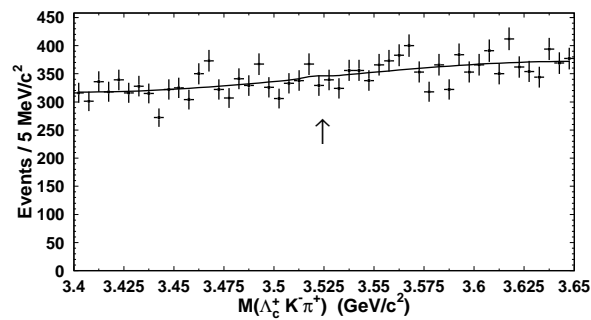}
\caption{ The $M(\Lambda_c^+ K^-\pi^+)$ distribution
near 3520 MeV/$c^2$ (indicated by an arrow), 
the mass of a possible doubly charmed baryon candidate~\cite{selex}.  
}
\label{3520} 
\end{figure}

The SELEX Collaboration reported the observation of a doubly charmed baryon 
with a mass of $3520$~MeV/$c^2$ in the $\Lambda_c^+ K^-\pi^+$ final state~\cite{selex}.  
We extend the range of the $M(\Lambda_c^+ K^- \pi^+)$ search to include
the region surrounding $3520$~MeV/$c^2$ (Fig.~\ref{3520}).
To compare the yield to the inclusive production of $\Lambda_c^+$, we  
modify the momentum requirement 
$p^*>2.5$~GeV/$c$ only for the $\Lambda_c^+$ baryon. 
We find no evidence for a signal either at this mass or in a wide 
range around it. The overlaid curve in Fig.~\ref{3520}
  is the result of the fit.
To describe a possible signal 
we use a Gaussian resolution function with the width fixed to the signal 
MC value of 4.9 MeV/$c^2$.
The background is parameterized by a third-order polynomial function. 
From the fit, we obtain an upper limit of 69.1 events at 90$\%$ 
confidence level (C.L.) ~When the same selection criteria are applied for the inclusive $\Lambda_c^+$ ($p^*>2.5$~GeV/$c$) production, we reconstruct $(83.5\pm 1.4)\times 10^4$ $\Lambda_c^+$ decays. Taking into account the ratio of the
 total reconstruction efficiencies, 
we derive an upper limit on the ratio of production cross sections with 
$p^*(\Lambda_c^+)>2.5$~GeV/$c$,~  
$\sigma(\Xi_{cc}(3520)^+)
       \times{\cal B}(\Xi_{cc}(3520)^+\to\Lambda_c^+ K^-\pi^+)
    /      \sigma(\Lambda_c^+)
     < 1.5\times10^{-4}$~at~90$\%$~C.L. 
Recently, the BaBar collaboration has also performed an extensive search for
doubly charmed baryons. 
They set an upper limit of $2.7\times 10^{-4}$ at 95$\%$ C.L.~\cite{babar_dcb} for the same decay process taking the
account the efficiencies of the $p^*$ requirement.

 In conclusion, we report the first observation of two charged baryons
 $\Xi_{cx}(2980)^+$ and $\Xi_{cx}(3077)^+$ decaying into $\Lambda_c^+
 K^-\pi^+$.
 We also search for neutral isospin related partners in $\Lambda_c^+
 K_S^0\pi^-$
 final state and observe a signal for the $\Xi_{cx}(3077)^0$.
 The statistical significance of each of these signals is more than 5$\sigma$.
 The masses and widths of all the observed states are summarized in Table I.
 Taking into account the presence of $s$ and $c$ quarks in the final state
 and the observation of an isospin partner near
3077~MeV/$c^2$
 in the $\Lambda_c^+ K_S^0\pi^-$ final state, the most natural
 interpretations
 of these states are that they are excited charmed strange baryons, $\Xi_c$.
 In contrast to decays of known excited $\Xi_c$ states the observed baryons
 decay into separate charmed ($\Lambda_c^+$) and strange ($K$) hadrons.
 Further studies of the properties of the observed states are ongoing.
We have also searched for the doubly charmed baryon state at 3520~MeV/$c^2$ reported by the SELEX collaboration in
the
 $\Lambda_c^+ K^-\pi^+$ final state~\cite{selex}, and extract an upper
 limit on
 its production cross section relative to the inclusive $\Lambda_c^+$ yield.

We thank the KEKB group for excellent operation of the
accelerator, the KEK cryogenics group for efficient solenoid
operations, and the KEK computer group and
the NII for valuable computing and Super-SINET network
support.  We acknowledge support from MEXT and JSPS (Japan);
ARC and DEST (Australia); NSFC (contract No.~10175071,
China); DST (India); the BK21 program of MOEHRD, and the
CHEP SRC and BR (grant No. R01-2005-000-10089-0) programs of
KOSEF (Korea); KBN (contract No.~2P03B 01324, Poland); MIST
(Russia); MHEST (Slovenia);  SNSF (Switzerland); NSC and MOE
(Taiwan); and DOE (USA).


\begin{thebibliography}{99}
\bibitem{lmc_sidebands} For the $\Lambda_c^+$ sidebands we use the $p K^-\pi^+$ combinations 
taken from $\pm 10$~MeV/$c^2$ mass intervals centered at $\pm 5~\sigma$ away of the $\Lambda_c^+$ mass.
\end{thebibliography}
\end{document}